\documentclass[seceq]{ptptex}

\usepackage{bbm}
\usepackage{amssymb}  
\usepackage{exscale}
\usepackage{textcomp}
\usepackage{amsmath, amsfonts}
\usepackage{epsfig}
\usepackage{graphicx}
\newcommand{\be}{\begin{equation}}
\newcommand{\ee}{\end{equation}}
\newcommand{\bea}{\begin{eqnarray}}
\newcommand{\eea}{\end{eqnarray}}

\title{
Holographic chiral currents in a magnetic field %
}

\author{
Anton \textsc{Rebhan}, Andreas \textsc{Schmitt}, Stefan \textsc{Stricker}%
}

\inst{
Institut f\"{u}r Theoretische Physik, Technische Universit\"{a}t Wien, 1040 Vienna, Austria
}

\abst{
In the presence of a quark chemical potential, a magnetic field induces an axial current in the direction of the magnetic field. We compute 
this current in the Sakai-Sugimoto model, a holographic model which, in a certain limit, is dual to large-$N_c$ QCD. We also compute the 
analogous vector current, for which an axial chemical potential is formally introduced. This vector current can potentially be observed via
charge separation in heavy-ion collisions. After implementing the correct axial anomaly in the Sakai-Sugimoto model we find an axial current 
in accordance with previous studies and a vanishing vector current, in apparent contrast to previous weak-coupling calculations.
}

\begin{document}

\maketitle

\section{Introduction}

The Sakai-Sugimoto model \cite{Sakai:2004cn,Sakai:2005yt} is a realization of the gauge/gravity duality \cite{Maldacena:1997re} 
which provides a top-down approach to a holographic model of QCD. While at present no gravity dual to QCD is known, the Sakai-Sugimoto model 
exhibits some of the most important properties of QCD, most notably confinement and chiral symmetry breaking. Confinement 
is realized within a background geometry of $N_c$ D4-branes, extending in the 3+1 dimensions of the field theory and an extra dimension $x^4$
\cite{Witten:1998zw}. This extra dimension is compactified on a circle with radius $M_{\rm KK}^{-1}$, 
where the Kaluza-Klein mass $M_{\rm KK}$ is a parameter of the model and sets the scale for the mass of unwanted adjoint scalars and fermions. 
Two solutions for the background geometry then account for confined and deconfined phases with a first-order phase transition 
between them at a critical temperature $T_c=M_{\rm KK}/(2\pi)$.

Fundamental chiral (and massless) fermions are described by introducing $N_f$ D8- and $\overline{\rm D8}$-branes (here and in most 
applications treated as probe branes), 
separated in the extra dimension \cite{Sakai:2004cn}. The gauge symmetries on these branes account for the global chiral symmetry 
$SU(N_f)_L\times SU(N_f)_R$ in the associated field theory. Two qualitatively different embeddings of the flavor branes are interpreted
as chirally symmetric (D8- and $\overline{\rm D8}$-branes separated) and chirally broken (D8- and $\overline{\rm D8}$-branes connected)
phases. 
In the version of the model used here, chiral symmetry breaking and confinement are equivalent, i.e., the ground state is chirally 
broken if and only if it is confined (by choosing a sufficiently small distance between D8- and $\overline{\rm D8}$-branes in the extra 
dimension one finds a chirally broken, deconfined phase; however, from the geometry of the model it is obvious that there cannot be
a chirally restored, confined phase).  

Here we are interested in the evaluation of the model in the presence of a quark chemical potential and a magnetic field. It has 
been shown that a magnetic field gives rise to baryon number even in the chirally broken phase \cite{Bergman:2008qv,Thompson:2008qw,Rebhan:2008ur}
(the resulting baryon density being homogeneous, in contrast to instanton-like baryons in the 
Sakai-Sugimoto model \cite{Hata:2007mb}). In our context, two physical situations are of special interest. First, large magnetic fields
can be found in the interior of compact stars (surface magnetic fields of magnetars being of the order of $B\lesssim 10^{15}\, {\rm G}$), where
also the quark chemical potential is large, $\mu \sim (400 - 500)\, {\rm MeV}$. In this case, we may expect an axial current in the direction of the
magnetic field \cite{Metlitski:2005pr,Newman:2005as}. 
Second, large magnetic fields can be temporarily created in 
noncentral heavy-ion collisions, reaching values of the order of $B\sim 10^{17}\,{\rm G}$. In the latter case, 
the interplay of the magnetic field with a nonvanishing chirality $N_5$, i.e., a nonzero difference between the number of left- and right-handed
fermions, may lead to the so-called chiral magnetic effect \cite{Kharzeev:2007jp,Fukushima:2008xe}. In the chiral magnetic effect, an
electric current is generated in the direction of the magnetic field, which possibly is responsible for the charge separation that has been observed
at the Relativistic Heavy-Ion Collider (RHIC) \cite{:2009uh}. The nonzero $N_5$ is provided by the QCD axial anomaly which relates certain gluon
configurations to a change in $N_5$, and thus to a nonzero current on an event-by-event basis. Here and in the original weak-coupling
calculations, the gluon field fluctuations are not taken into account explicitly, a nonzero $N_5$ is rather described by introducing 
an axial chemical potential $\mu_5$. This description has to be taken with some care since $N_5$ is not a conserved quantity and thus
strictly speaking there is no such thing as an axial chemical potential in the thermodynamic sense.

In summary, we shall be interested in the axial current in the presence of a vector chemical potential and the vector current in the presence
of an axial chemical potential. More details about the following calculations can be found in Ref.\ \cite{Rebhan:2009vc}.

\section{Currents and anomalies}
\label{sec:currents}

We start from the (Euclidean) action of the Sakai-Sugimoto model for one quark flavor \cite{Sakai:2005yt},
\be
S = S_{\rm YM} + S_{\rm CS} \, ,
\ee
with Yang-Mills (YM) and Chern-Simons (CS) contributions
\begin{subequations} \label{SYMCS}
\bea
S_{\rm YM} &=& \kappa M_{\rm KK}^2 \int d^4 x\int_{-\infty}^{\infty} dz \, 
\left[k(z)F_{z\mu}F^{z\mu}+\frac{h(z)}{2M_{\rm KK}^2}F_{\mu\nu}F^{\mu\nu}\right] \, , \label{SYM}\\
S_{\rm CS} &=& \frac{N_c}{24\pi^2} \int d^4 x\int_{-\infty}^{\infty} dz\, A_\mu F_{z\nu}F_{\rho\sigma}\epsilon^{\mu\nu\rho\sigma} \, , \label{SCS}
\eea
\end{subequations}
with Greek indices running over $\mu,\nu,\ldots = 0,1,2,3$. 
Our convention for the epsilon tensor is $\epsilon_{0123}=+1$. 
We have used the metric functions  
\be
k(z)\equiv 1+z^2 \, , \qquad h(z)\equiv (1+z^2)^{-1/3} \, , 
\ee
and the dimensionless constant
\be \label{kappa}
\kappa \equiv \frac{\lambda N_c}{216\pi^3} \, ,
\ee
where $\lambda$ is the 't Hooft coupling. The integration is over 
space-time $(\tau,{\bf x})$ and the holographic (dimensionless) coordinate $z$. In the confined phase, $z$ extends from the 
left-handed boundary ($z=+\infty$) over
the tip of the cigar-shaped $(x^4,z)$ subspace ($z=0$) to the right-handed boundary ($z=-\infty$). (In the deconfined phase, 
the integral over connected D8 and $\overline{\rm D8}$ branes is replaced by two separate integrals over disconnected D8 and 
$\overline{\rm D8}$ branes; in this
section we restrict ourselves to the confined phase, all arguments are analogous in the deconfined phase.) 

The equations of motion 
are
\begin{subequations} \label{eom}
\bea
\kappa M_{\rm KK}^2\,
\partial_z[k(z)F^{z\mu}]+\kappa
h(z)\partial_\nu F^{\nu\mu} &=&\frac{N_c}{16\pi^2}F_{z\nu}F_{\rho\sigma}
\epsilon^{\mu\nu\rho\sigma} \, , \label{eq1}\\
\kappa M_{\rm KK}^2\,
\partial_\mu[k(z)F^{z\mu}] &=&\frac{N_c}{64\pi^2}F_{\mu\nu}F_{\rho\sigma}\epsilon^{\mu\nu\rho\sigma} \, , \label{eq2}
\eea
\end{subequations}
where the second equation 
is obtained from varying $A_z$ (which, in our gauge choice, has already been set to zero in the action (\ref{SYMCS})).

Next we introduce the chiral
currents. They are defined through the variation of the on-shell action with respect to the boundary values,
\be \label{currents}
{\cal J}^\mu_{L/R} \equiv -\frac{\delta S}{\delta A_\mu(x,z=\pm\infty)}= 
\mp\left(2\kappa M_{\rm KK}^2k(z)F^{z\mu} -\frac{N_c}{24\pi^2}\epsilon^{\mu\nu\rho\sigma}
A_\nu F_{\rho\sigma}\right)_{z=\pm\infty} \, ,
\ee
where the first (second) term is the YM (CS) contribution. 
It is only the YM part of the current, i.e., the first term in eq.\ (\ref{currents}), which appears in the asymptotic
expansion of the gauge fields \cite{Rebhan:2008ur}: from the definition (\ref{currents}) and with $k(z)=1+z^2$ we find
\bea \label{expandYM}
A_\mu(x,z) &=& A_\mu(x,z=\pm\infty)\pm \frac{{\cal J}_{\mu,{\rm YM}}^{L/R}}{2\kappa M_{\rm KK}^2}\frac{1}{z} 
+ {\cal O}\left(\frac{1}{z^2}\right) \, .
\eea
One can also confirm this relation from our explicit results below. 

The divergence of the currents (\ref{currents}) can be easily computed with the help of the equation of motion for $A_z$ (\ref{eq2}). 
With the left- and right-handed field strengths $F_{\mu\nu}^{L/R}(x) \equiv F_{\mu\nu}(x,z=\pm\infty)$, 
the corresponding dual field strength tensors $\widetilde{F}^{\mu\nu}_{L/R}=\frac12\,F_{\rho\sigma}^{L/R}\epsilon^{\mu\nu\rho\sigma}$, 
the vector and axial currents ${\cal J}^\mu \equiv {\cal J}^\mu_R+{\cal J}^\mu_L$, ${\cal J}^\mu_5 \equiv {\cal J}^\mu_R-{\cal J}^\mu_L$, 
and the vector and axial field strengths introduced as $F_{\mu\nu}^R=F_{\mu\nu}^V+F_{\mu\nu}^A$, $F_{\mu\nu}^L=F_{\mu\nu}^V-F_{\mu\nu}^A$, 
we obtain the vector and axial anomalies
\begin{subequations}
\bea
\partial_\mu{\cal J}^\mu &=& \frac{N_c}{12\pi^2} F_{\mu\nu}^V\widetilde{F}^{\mu\nu}_A \, , \\
\partial_\mu{\cal J}^\mu_5 &=& \frac{N_c}{24\pi^2} \left(F_{\mu\nu}^V\widetilde{F}^{\mu\nu}_V+F_{\mu\nu}^A\widetilde{F}^{\mu\nu}_A\right) \, .
\label{consanomaly1b}
\eea
\end{subequations}
The coefficients on the right-hand side 
(which receive contributions from both
the YM and CS parts of the currents) are in accordance with the standard
field-theoretic results for $N_c$ chiral fermionic
degrees of freedom coupled to left and right chiral gauge fields 
\cite{Bardeen:1969md}. 
The above form of the anomaly, which is symmetric in vector and axial-vector gauge fields, is called {\it consistent} anomaly.
If left- and 
right-handed Weyl spinors are treated separately, 
this form of the anomaly arises unambiguously. 
In QED, however, we must require that 
the vector current be strictly conserved, 
even in the presence of axial field strengths.
As was first discussed by Bardeen \cite{Bardeen:1969md}, this
can be achieved by the introduction of a counterterm that
mixes left- and right-handed gauge fields.
Having even parity, Bardeen's counterterm is uniquely given by \cite{Hill:2006ei}   
\be\label{DeltaS}
\Delta S = c\int d^4x (A_\mu^L A_\nu^R F_{\rho\sigma}^L+A_\mu^L A_\nu^R F_{\rho\sigma}^R)\epsilon^{\mu\nu\rho\sigma} \, ,
\ee
where $c$ is a constant determined by requiring a strictly conserved vector current. Because this expression can be naturally written as a (metric-independent) 
integral over a
hypersurface at $|z|=\Lambda\to\infty$ with left- and right-handed fields
concentrated at the respective brane locations, $\Delta S$ can
actually be interpreted as a (finite) counterterm in holographic renormalization.
In particular, it does not change the equations of motion.

The contribution of Bardeen's counterterm to the chiral currents is 
\be \label{DeltaJ}
\Delta {\cal J}^\mu_{L/R} = \mp c \left(A_\nu^{R/L}F_{\rho\sigma}^{R/L}-A_\nu^{L/R}F_{\rho\sigma}^{R/L}+2A_\nu^{R/L}F_{\rho\sigma}^{L/R}\right)
\epsilon^{\mu\nu\rho\sigma} \, .
\ee
Denoting the renormalized left- and right-handed currents by $\bar{\cal J}^\mu_{L/R} \equiv {\cal J}^\mu_{L/R} + \Delta{\cal J}^\mu_{L/R}$,
and $\bar{\cal J}^\mu \equiv \bar{\cal J}^\mu_R+\bar{\cal J}^\mu_L$, $\bar{\cal J}^\mu_5 \equiv \bar{\cal J}^\mu_R-\bar{\cal J}^\mu_L$, 
we find that $c=N_c/(48\pi^2)$ is required to obtain the {\em covariant} anomaly
\begin{subequations}\label{covanomaly}
\bea
\partial_\mu \bar{\cal J}^\mu &=& 0  \, , \label{covanomaly1} \\
\partial_\mu \bar{\cal J}^\mu_5 &=& \frac{N_c}{8\pi^2} F_{\mu\nu}^V \widetilde{F}^{\mu\nu}_V
+ \frac{N_c}{24\pi^2} F_{\mu\nu}^A \widetilde{F}^{\mu\nu}_A \, . \label{covanomaly2} 
\eea
\end{subequations}
The prefactor in front of the first term in the axial anomaly now has changed from $N_c/(24\pi^2)$ in 
eq.\ (\ref{consanomaly1b}) to $N_c/(8\pi^2)$, which is the well-known result for the
Adler-Bell-Jackiw anomaly for QED \cite{Bell:1969ts,Adler:1969gk}
and which is essential for getting the correct pion decay rate $\pi^0\to 2\gamma$. 
The necessity of adding the counterterm (\ref{DeltaS}) to the Sakai-Sugimoto
model is in fact completely analogous to the well-known procedure 
in chiral models where a Wess-Zumino-Witten term accounts for the anomaly \cite{Kaymakcalan:1983qq}. 

In the literature sometimes the coefficient of the subleading term in the asymptotic
behavior of $A_\mu(z)$ and thus the YM part of the current (see eq.~(\ref{expandYM}))
is identified with the full current \cite{Yee:2009vw,Son:2009tf}. Using this
identification, it has also been assumed that the equation of motion for $A_z$ (\ref{eq2}) represents the anomaly equation \cite{Lifschytz:2009si}.
Indeed, from eq.\ (\ref{eq2}) one obtains 
\begin{subequations}\label{YMcurranomalies}
\bea
\partial_\mu{\cal J}^\mu_{\rm YM} &=& \frac{N_c}{4\pi^2} F_{\mu\nu}^V\widetilde{F}^{\mu\nu}_A \, , \\
\partial_\mu{\cal J}^\mu_{\rm YM,5} &=& \frac{N_c}{8\pi^2} \left(F_{\mu\nu}^V\widetilde{F}^{\mu\nu}_V+F_{\mu\nu}^A\widetilde{F}^{\mu\nu}_A\right) \, ,
\eea
\end{subequations}
and this does contain the same coefficient in front of $F_{\mu\nu}^V\widetilde{F}^{\mu\nu}_V$ as the full covariant anomaly (\ref{covanomaly}).
However, it differs from the latter in the presence of axial gauge fields. In particular, the vector current is then not strictly conserved. 
Even when this issue may be ignored, because all
axial vector field strengths are set to zero, it appears to be questionable
to keep only part of the full current (\ref{currents}).

\section{Solution to the equations of motion}

We introduce a homogeneous magnetic field $B$ in the spatial 3-direction via the gauge field component $A_1(x_2) = -x_2 B$, i.e., $A_1$ 
is a constant with respect to the holographic coordinate $z$, which is consistent with the equations of motion. 
Then, the only nontrivial gauge field components we need are $A_0(z)$ and $A_3(z)$, where $A_0$ is necessary to account for finite chemical 
potentials, and $A_3$ becomes nonzero through the equations of motion. (With only a little more effort, the equations of motion can also
be solved in the presence of an additional electric field\cite{Rebhan:2009vc}. This is for instance useful to check the axial anomaly 
explicitly, but for simplicity we restrict ourselves to a magnetic field which is the physically relevant field in our context.) 

\subsection{Chirally broken phase}

Within this ansatz, the equations of motion read 
\bea \label{E123}
\partial_z(k\partial_z A_0) &=& 2\beta \partial_z A_3 \, , \qquad
\partial_z(k\partial_z A_3) = 2\beta \partial_z A_0 \, , 
\eea
with the dimensionless magnetic field $\beta\equiv \alpha B/M_{\rm KK}^2$
where $\alpha\equiv 27\pi/(2\lambda)$. The general solutions to Eqs.\ (\ref{E123}) are
\begin{subequations} \label{gaugefields1}
\bea
A_0(z) &=& \mu-\mu_{5}\frac{\sinh(2\beta\arctan z)}{\sinh\beta\pi} 
-\jmath \left[\frac{\cosh(2\beta\arctan z)}{\sinh\beta\pi} -\coth\beta\pi\right] \, , \\
A_3(z) &=& -\mu_{5}\left[\frac{\cosh(2\beta\arctan z)}{\sinh\beta\pi}-\coth\beta\pi\right] 
-\jmath\, \frac{\sinh(2\beta\arctan z)}{\sinh\beta\pi}  \, .
\eea
\end{subequations}
Here we have identified the boundary values of the temporal components of the gauge fields with the chemical potentials, 
$A_0(z=\pm\infty)=\mu_{L,R}$, 
and the vector and axial chemical potentials are $\mu\equiv (\mu_R+\mu_L)/2$ and $\mu_5\equiv(\mu_R-\mu_L)/2$. The boundary values of the 
spatial components of the gauge fields are $A_3(z=\pm\infty)=\mp\jmath$. 
 The axial supercurrent $\jmath$, which has to be determined from minimizing the free energy, describes a rotation of the chiral condensate 
in the space spanned by the scalar and pseudoscalar mesons \cite{Rebhan:2008ur} and thus may be related to the recently discussed quarkyonic 
chiral spirals \cite{Kojo:2009ha}.

\subsection{Chirally restored phase}

In the case of disconnected D8 and $\overline{\rm D8}$ branes, i.e., in the deconfined, chirally restored phase, we have two 
sets of equations of motion, one for the left-handed fields $A_\mu^L(z)$ and one for the right-handed fields $A_\mu^R(z)$ with $z\in [0,\infty]$,
\bea \label{E123sym}
\partial_z(k_0\partial_z A_0^{L/R}) &=& \pm \frac{2\beta}{\theta^3} \partial_z A_3^{L/R} \, , \qquad 
\partial_z(k_3\partial_z A_3^{L/R}) = \pm \frac{2\beta}{\theta^3} \partial_z A_0^{L/R} \, . 
\eea
Here  $\theta \equiv 2\pi T/M_{\rm KK}$ is the dimensionless temperature, and $k_0(z)\equiv \frac{(1+z^2)^{3/2}}{z}$, 
$k_3(z)\equiv z(1+z^2)^{1/2}$
are different metric functions for temporal and spatial components of the gauge fields. The general solution to Eqs.\ (\ref{E123sym}) is
\begin{subequations} \label{gaugefields2}
\bea
A_0^{L/R}(z) &=& (\mu\mp\mu_{5})\left[p(z)-\frac{p_0}{q_0}\,q(z)\right] \, , \\
A_3^{L/R}(z) &=& \pm\frac{\mu\mp\mu_{5}}{2\beta/\theta^3}
\left[k_0\partial_zp-\frac{p_0}{q_0}(1+k_0\partial_zq)\right]
\, ,
\eea
\end{subequations}
where $p_0\equiv p(0)$, $q_0\equiv q(0)$, and 
\begin{subequations} \label{pq}
\bea
p(z) &=& {}_2F_1\left[-\frac{\sqrt{1-16\beta'^2}+1}{4},\frac{\sqrt{1-16\beta'^2}-1}{4},\frac{1}{2},
\frac{1}{1+z^2}\right] \, , \\
q(z) &=& \frac{1}{\sqrt{1+z^2}}\, {}_2F_1\left[-\frac{\sqrt{1-16\beta'^2}-1}{4},\frac{\sqrt{1-16\beta'^2}+1}{4},\frac{3}{2},
\frac{1}{1+z^2}\right] \, ,
\eea
\end{subequations}
with the abbreviation $\beta'\equiv \beta/\theta^3$. Again, the boundary values of the temporal components 
of the gauge fields are identified with the chemical potentials, $A_0^{L/R}(\infty) = \mu_{L/R}$. Following Ref.\ \cite{Horigome:2006xu}, 
these components vanish at the horizon, $A_0^{L/R}(0) = 0$ (see however Refs.\ \cite{Yeetalk,Gynther:2010ed}). 
The spatial components vanish at the holographic boundary, but acquire a finite value at the horizon,
$A_3^{L/R}(0)=\mp\frac{\mu_{L/R}}{2\beta/\theta^3}\frac{p_0}{q_0}$.

\section{Results and discussion}

To compute the axial and vector currents in the direction of the magnetic field we use the solutions to the equations of 
motion, Eqs.\ (\ref{gaugefields1}) and (\ref{gaugefields2}), and insert them into the chiral currents, which are defined through 
Eq.\ (\ref{currents}) and the contribution from Bardeen's counterterm (\ref{DeltaJ}). The results are summarized in Table \ref{tab:summary}.

Let us first discuss the full renormalized currents $\bar{\mathcal J} = \mathcal J_{\rm YM} + \mathcal J_{\rm CS}+\Delta \mathcal J$ 
which have been obtained under the constraint of the 
correct (covariant) QED anomaly. For the axial current we find that the counterterm $\Delta \mathcal J$ exactly cancels the CS part. 
In the chirally symmetric phase, this yields exactly the expected topological result ${\cal J}_5=\frac{\mu B N_c}{2\pi^2}$
\cite{Metlitski:2005pr,Newman:2005as}.
In particular, the dimensionful parameter of the model, $M_{\rm KK}$, drops out of this universal result.
In the chirally broken phase, the axial current is smaller and given by a complicated function of $B$. 

The most striking of our results is that for both phases the renormalized vector current is zero for all magnetic fields.
One might have expected that, in the deconfined phase, this current should reproduce the known result for the chiral magnetic 
effect \cite{Fukushima:2008xe}, ${\cal J}=\frac{\mu_5 B N_c}{2\pi^2}$, because this result can be derived from the anomaly 
only \cite{Nielsen:1983rb} and we have explicitly made sure that our model reproduces the correct anomaly. The question is thus whether our result
can indeed be interpreted as the current responsible for the chiral magnetic effect. We shall comment on possible issues related to this question 
below. Our result of a vanishing current in the 
confined phase, however, seems less puzzling. The usual explanation of the chiral
magnetic effect, using a quasiparticle picture (which is not guaranteed to hold in our strong-coupling approach), relies on 
individual, electrically charged, massless quarks which move in different directions according to their chirality. A suppression
of the effect may thus indeed be expected in the confined, chirally broken phase \cite{Fukushima:2008xe,Fu:2010rs}.

\newcommand\V{F_V\widetilde F_V }
\newcommand\A{F_A\widetilde F_A }
\newcommand\VA{F_V\widetilde F_A }


\renewcommand{\arraystretch}{1.2}
\begin{table}[t]
\begin{tabular}{|c|c||c|c|c|c|} 
\hline
&  & $\mathcal J_{\rm YM}$ & $\mathcal J_{\rm YM} + \mathcal J_{\rm CS}$ & $\mathcal J_{\rm YM} + \mathcal J_{\rm CS}+\Delta \mathcal J$ \\
\hline\hline
 & anomaly & ``semi-covariant'': & consistent: & \underline{covariant:}  \\
\cline{2-5}
\rule[-1.5ex]{0em}{4.5ex} 
& $\partial_\mu \mathcal J^\mu_5 / \frac{N_c}{24\pi^2}$ & 
 \underline{$3 \V$}$+3\A$ & $\V+\A$ & \underline{$3\V+\A$} \\[0.5ex]
\cline{2-5}
\rule[-1.5ex]{0em}{4.5ex} 
& $\partial_\mu \mathcal J^\mu / \frac{N_c}{24\pi^2}$ & 
 $6\VA$ & $2\VA$ & \underline{$0$}  \\[0.5ex]
\hline\hline
\rule[-1.5ex]{0em}{4.5ex} 
$\;\;T>T_c\;\;$ & $\mathcal J_5 / \frac{\mu B N_c}{2\pi^2}$ & 
 1 & $\frac23$ & 1 \\[1ex]
\cline{2-5}
\rule[-1.5ex]{0em}{4.5ex} 
& $\mathcal J / \frac{\mu_5 B N_c}{2\pi^2}$ & 
 1 & $\frac23$ & 0  \\[0.5ex]
\hline\hline
\rule[-1.5ex]{0em}{4.5ex} 
$\;\;T<T_c\;\;$ & $\mathcal J_5 / \frac{\mu B N_c}{2\pi^2}$ & 
 ${\frac{\beta\coth\beta\pi}{2\rho(\beta)}}$  & ${\frac{\beta\coth\beta\pi}{2\rho(\beta)}-\frac{1}{3}}$ & 
${\frac{\beta\coth\beta\pi}{2\rho(\beta)}}$ \\[1ex]
\cline{2-5}
\rule[-1.5ex]{0em}{4.5ex} 
 & $\mathcal J / \frac{\mu_5 B N_c}{2\pi^2}$ & 
 1  & $\frac23$ & 0 \\[1ex]
\hline
\end{tabular}
\caption{Different (parts of the) axial and vector currents ${\cal J}_5$ and ${\cal J}$ -- normalized to the weak-coupling results
\cite{Metlitski:2005pr,Fukushima:2008xe} -- 
in the direction of the magnetic field
$B$ in the deconfined, chirally symmetric phase ($T>T_c$) and in the confined, chirally broken phase ($T<T_c$). We have abbreviated
$\rho(\beta)\equiv \beta\coth\beta\pi+\frac{\pi\beta^2}{2\sinh^2\beta\pi}$ with the dimensionless magnetic field 
$\beta=\alpha B/M_{\rm KK}$.
}\label{tab:summary}
\end{table}

In Table \ref{tab:summary} we also show the results which are obtained from the YM part only. In the case of the axial currents, there is no 
difference to the renormalized currents. For the vector currents, however, we now recover the weak-coupling (deconfined) result 
for both confined and deconfined phases. Therefore, now the deconfined current is as expected while the lack of suppression in the 
confined phase comes as a surprise. 

There are several problems in the current approach. First, upon computing the free energy explicitly and then taking the derivative with respect 
to the appropriate source, the currents turn out to be different from the straightforward definition via the gauge/gravity correspondence
(used for the results in Table \ref{tab:summary}). This disturbing discrepancy arises for nonvanishing background magnetic fields and
can be attributed to boundary terms at spatial infinity. A previously suggested fix of this problem by modifying the action \cite{Bergman:2008qv} 
seems to be not acceptable for our purpose 
because it entirely eliminates the axial anomaly from the correspondingly modified currents \cite{Rebhan:2009vc}. 
Second, the introduction of a $\mu_5$ may be problematic. Since a chemical potential must be associated to a conserved charge, it has been argued 
that a physically meaningful vector current can be obtained only after introducing a conserved (not anomalous) charge $N_5$ 
\cite{Alekseev:1998ds,Rubakov:2010qi}. This, on the other hand, corresponds to a gauge {\it variant} axial charge density,
which precludes a generalization to inhomogeneous situations \cite{Gynther:2010ed}.

In a simple holographic model not unlike the Sakai-Sugimoto model, and in a setup that corresponds to our deconfined phase, the vector current 
has recently been found to coincide with the weak-coupling result \cite{Gynther:2010ed}. In this study, which uses linear response theory (i.e., 
only deals with an infinitesimally small magnetic field), it has been pointed out that it is important to 
distinguish the source for the quark density (= boundary value of the gauge field) from the chemical potential (= potential difference between 
boundary and horizon). With this distinction the sources can be set to zero,  and consequently the CS contributions to the currents vanish
trivially. It remains to be seen whether this concept can be applied to the present model, in particular to the case of a finite 
background magnetic field \cite{prepare}. Moreover, the situation in the confined phase seems profoundly different, since in the case of connected 
branes it is unclear how to define the chemical potential other than through the boundary value. Maybe this difference is not surprising
because it is the deconfined phase in which the results of the renormalized currents seem puzzling and the YM currents seem to agree with 
physical expectations. In the confined phase it is the other way around.

\section*{Acknowledgements}
A.S.\ thanks the organizers of the ``New Frontiers of QCD 2010'' program for the invitation and a stimulating atmosphere at the Yukawa 
Institute for Theoretical Physics in Kyoto. This work has been supported by FWF project no.\ P19958.

%

\end{document}